\documentclass{amsart}
\usepackage{amssymb, array, amscd}
\usepackage{amsmath, amsfonts,amsthm,graphicx}
\theoremstyle{plain}

\numberwithin{equation}{section}

\newcommand{\Z}{\mathbb{Z}}

\begin{document}

\title{ A probabilistic regulatory network  for the human immune  system.}

\author{Mar\'{\i}a A. Avi\~no-Diaz }
\address{Department of Mathematics, UPR-Cayey , PR 00736}
\email{mavino@uprr.pr}
\begin{abstract}
In this paper we made a review of some papers about probabilistic
regulatory networks (PRN), in particular we introduce our concept of
homomorphisms of PRN  with an example  of projection of a regulatory
network to a smaller one. We apply the model PRN ( or Probabilistic
Boolean Network) to the immune system, the PRN works with two
functions. The model called  "The B/T-cells interaction" is Boolean,
so we are really working with a Probabilistic Boolean Network. Using
Markov Chains we determine the state of equilibrium of the immune
response.\end{abstract}
 \maketitle
 \section{Introduction}
 The biological process can be modeled using different class of models,
 but in most of the applications differential equation models  have been selected
 because the entities can have more than two values. Here we use a probabilistic
 regulatory network model, that it is a  discrete model with only a
 finite number of state and activities. In particular, we describe  the dynamic
  of the immune response in humans, using the Boolean model called
  B/T-cells \cite{KUT}, and we added probabilities and use the dynamic model of
  Probabilistic Boolean Network, \cite{S1,S2,S3, SDZ,S4}.

The immune system is  separated  by functionality in two parts:
recognition and ''effector'' functions. A complex immune system has
cells and molecules that give us a basic defense against bacteria,
viruses, fungi, and other pathogenic agents. Possibly, it has tens
or hundreds of different types of regulatory and effector molecules.
So, an important role in the study of this class of system plays the
reduction of networks, for that reason we introduce the concept of
projection of one net to another smaller for the future
applications. \emph{A variety of cell types compose the immune
system, the most important are the lymphocytes. These cells are
created in the bone marrow, along with all of the other blood cells,
and are transported throughout the body via the blood stream.
Lymphocytes spend considerable time resident in lymphoid organs,
such as the bone marrow, the thymus, the spleen, and lymph nodes.
Lymphocytes are subdivided in two classes: B-cells and T-cells}, see
Perelson \cite{P}. B lymphocytes secrete antibodies, and the main
function of T-cells is the interacting with other cells. Helper T
cells act through the secretion of lymphokines, they made possible
to transform the B-cells into an antibody-secreting state. Helper
T-cells are the cells that are predominantly infected by the human
immunodeficiency virus, and plays a major role in AIDS. Cytotoxic
T-cells, are responsible for killing virally infected cells and
cells that look like  abnormal, such as some tumor cells.

In this paper we study the dynamic of the B/T-cell model giving by
Kaufman, Urbain, and Thomas, n 1985. In this model they use
functions, in which  a boolean parameter appears, and they obtained
 the steady state of the system, in our case we describe in a more
complex way the immune system using probabilities for the two
possible functions  of activity. Our model can be changed for
another more complex if we consider a three values model instead of
a Boolean model

\section{Preliminaries and projection}
In this section we introduce the mathematical background of the
model Probabilistic Regulatory Network, and the concept of
projection using and example. For the complete mathematical
background we suggest to see, \cite{A1,G,BL}. Here, we give a method
that permit us to build regulatory networks with probabilities
assigned to its functions. We use an algorithm for  understanding
the concept of Probabilistic Regulatory Network.

\subsection{Algorithm}\label{algorithm1}
Input:

1.  $n=$ number of entities in the network under studying,
for example 100 genes, and the set of values for each entity, that we denote by $k_a$. \\
2. A set of  relations $\{m_{a,b}\}$ taking $1$ if the entity $a$ is related to the entity  $b$, and $0$  otherwise.\\
3. A set of finite families of states in the network which gives the
time series data for one, two or more update functions,
 $A_1=\{(a^1_{i,1},\ldots,a^2_{i,n-1}, a^1_{i,n})|1\leq i\leq m_x \},$
  $\ldots$, and  $A_s=\{(a^s_{i,1},\ldots,a^s_{i,n-1}, a^s_{i,n})|1\leq i\leq
  m_s \}$.\\
4. A set of values $C=\{c_1,\ldots, c_s\}$ with $s$ probabilities
obtained in some way by the experiment or by the time series data.
That is $c_1+\cdots+c_s=1,$ and  $c_i\in[0,1]$.
\begin{itemize}
\item[(Alm1)] Creation the low level graph $\Gamma$:
\begin{itemize}
\item[1.]  $n=3$ , and  $m_{ 12}=1$ ,  $m_ {23}=1$ , y $m_{13}=0$,
 then our net is very simple and it is the following:
\[\hspace{.3in} \Gamma \ \hspace{.3in} \ \begin{array}{ccc}
&  & \bullet \ 3\\
&  & \mid \\
 1\ \bullet & {\overline{\hspace{.2in}}} &\bullet \ 2
\end{array}, \hspace{.3in}\ \]
\end{itemize}
\item [(Alm2)] We define the set where the functions are acting, in
our case, they are boolean, that is $k_1=\{0,1\}=k_2=k_3=\Z_2$,
considering
 $k= k_1 \times  k_2 \times  k_3=\{0,1\}^3$. The two sequential states are
 the
 following:

Time series data 1 $A_1=\{(0,1,0),(1,1,1),(1,1,0),(1,1,1)\}$.

Time series data 2  $A_2=\{(0,1,0),(0,1,1),(0,1,0),(0,1,1)\}$.\\

We obtain two different update functions:
\[f_1(x_1,x_2,x_3)=(1,x_2,x_2(x_3+1)),\ and \ f_2(x_1,x_2,x_3)=(x_1,x_2,x_2(x_3+1)).\]

\item [(Alm3)] We assign the following  probabilities to each update
function: $2/3$ to the function $f_1$, and $1/3$ for  the function
$f_2$.
\item[(Alm4)] We construct the high level digraph, that is the
following in this example
\end{itemize}
Output:
\[ States \ Space \]
\[\begin{array}{ccc}
\overset{1}{\circlearrowright}\underbrace{\overbrace{(1,0,0)}}&
\overset{1}{\leftarrow}&
\underbrace{\overbrace{(1,0,1)}}\\
^{2/3} \uparrow & ^{2/3} \nwarrow&\\
\underbrace{\overbrace{(0,0,1)}}&\overset{1/3}{\rightarrow}&
\underbrace{\overbrace{(0,0,0)}}\overset{1/3}{\circlearrowleft}\\
\end{array} \ \hspace{.3in} \begin{array}{ccc}
\underbrace{\overbrace{(0,1,0)}}&\overset{1/3}{\leftrightarrows}&
\underbrace{\overbrace{(0,1,1)}}\\
^{2/3}\downarrow & &^{2/3}\downarrow\\
\underbrace{\overbrace{(1,1,1)}}&\overset {1}{\leftrightarrows}&
\underbrace{\overbrace{(1,1,0)}}\\
\end{array}   \]

In order to study the dynamic we need the Transition Matrix of the
system, because we have two functions acting on the set of states.
So the dynamic will be study using Markov Chains. First we use the
following order for the states, but this is not the only possibility
\[\begin{array}{cccccccc}
(0,0,0)& (0,0,1) &(1,0,0)& (1,0,1)& (0,1,1)& (1,1,0)
&(0,1,0)&(1,1,1)\\
 1& 2& 3&  4&   5&   6&   7&   8\\
 \end{array}\]
The matrix is constructed in the following way $a_{(i,j)}$ is the
probability to have the arrow that it is going from  $i$ to  $j$,
then $a_{(2,3)}=p(f_2)=1/3$, but $a_{(3,3)}=p(f_2)+p(f_1)=1$ because
the two functions are going from $(1,0,0)$ to  $(1,0,0)$.
\[T=\left[\begin{array}{cccc|cccc}
0 &  2/3& 1/3& 0 &  0 &  0 &  0 &  0\\
 2/3 & 0 &  1/3& 0 &  0 &  0 &  0 &  0\\
 0 &  0 & 1 &  0& 0  &  0  &  0 &  0\\
  0 &  0 &  1 &  0& 0& 0 &  0 &  0\\
  \hline
   0 &  0 &  0 &  0& 0& 1/3& 2/3& 0\\
    0& 0&  0&   0&   0&   0&   0& 1\\
     0& 0&   0&   0&   2/3& 0& 0&   1/3\\
      0& 0& 0&   0&   0&   1&   0&   0\\
      \end{array}\right]\]
The dynamic of the systems is going to stationary states, or the
knowing by the equilibrium of the system. we have this information
with the iteration of the matrix, that is computing the power of the
transition matrix until we have the same vector in each arrow of the
matrix. In this very simple case we have two separated spaces, so
our matrix works with two sub matrices  of  $T$: $T_{11}$  and
$T_{22}$, in fact
\[T_{11}\rightarrow\left[\begin{array}{cccc}
0&0&1&0\\
0&0&1&0\\
0&0&1&0\\
0&0&1&0\\
\end{array}\right]\]
meanwhile for the other submatrix   $T_{22}$  we have:
\[T_{22}\rightarrow\left[\begin{array}{cccc}
0&0&0&1\\
0&1&0&0\\
0&1&0&0\\
0&0&0&1\\
\end{array}\right]\]
This induce that the equilibrium of the systems is the following
\[\pi=(0,0,.5,0,0,.25,0,.25)\]
That is, the system is going to the boolean  vector $(1,0,0)$ with a
probability of  .5, to the boolean vectors $(1,1,0),\ (1,1,1)$, with
probability of $.25$. We consider that, additionally some part of
this systems is going from  $(1,1,0)$ to $(1,1,1),$ and  from
$(1,1,1)$ to $(1,1,0)$ continuously.

\section{Projection and Reduction of networks}
Reduction of a network is our interest in this section, we use a
bigger net than the one in the last section. In particular the
mathematical concept that permit us to do that is the called
\emph{homomorphism}. In particular here we present a projection,
that it is an homomorphism which reduce the network to smaller and
have the very important property to have the similar state of
equilibrium, for mathematical background see \cite{A,AA}. For other
approach of the concept of homomorphism, that had applications to
dynamical networks, see \cite{DS,ID,MD}.
\begin{figure}
\includegraphics[height = 3in,width = 5in]{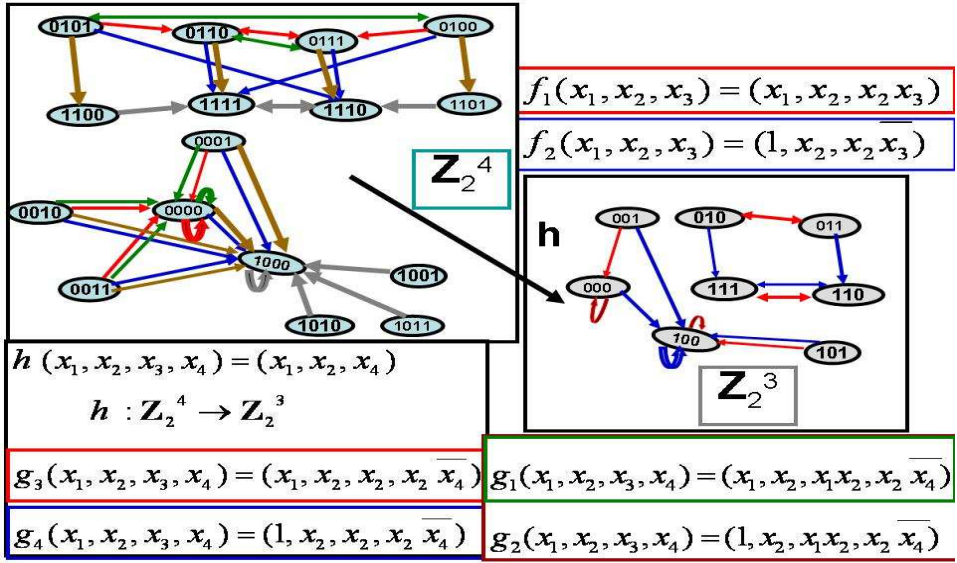}\caption{ Projection}
\end{figure}
We have the last example and now, the following network is defined
as follows: $\Delta$,
\[  \hspace{.4in} \Delta \
\ \hspace{.3in} \ \ \
\begin{array}{ccc}
 2\ \bullet &{\overline{\hspace{.2in}}} & \bullet \ 4\\
& \diagdown  & \mid \\
 1\ \bullet & {\overline{\hspace{.2in}}} & \bullet \ 3
\end{array}  \hspace{.2in}   \    \]
Each vertex has two values, that is our network  is boolean
$\Z_2=\{0,1\}$, then the functions act on the set  ${\Z_2}^4$, that
has  $16$ elements or states. The functions are the following:
\[\begin{array}{ll}
g_1(x_1,x_2,x_3,x_4)=(x_1,x_2,x_1x_2,x_2\overline{x_4}),&
g_2(x_1,x_2,x_3,x_4)=(1,x_2,x_1x_2,x_2\overline{x_4})\\
g_3(x_1,x_2,x_3,x_4)=(x_1,x_2,x_2,x_2\overline{x_4}),&
g_4(x_1,x_2,x_3,x_4)=( 1,x_2,x_2,x_2\overline{x_4})\\
\end{array}.\]
The homomorphism is the following:
\[h(x_1,x_2,x_3,x_4)=(x_1,x_2,x_4)\]
This function satisfies the following properties, that are called
commutative diagrams:\[\begin{CD}
{\Z_2}^4 @ > g_1 >>{\Z_2}^4 \\
 @V h VV  @V h VV  \\
{\Z_2}^3@ > f_1 >> {\Z_2}^3
\end{CD},  \  \begin{CD}
{\Z_2}^4 @ > g_3 >>{\Z_2}^4 \\
 @V h VV  @V h VV  \\
{\Z_2}^3@ > f_1 >> {\Z_2}^3
\end{CD}, \] \[  \begin{CD}
{\Z_2}^4@> g_2 >> {\Z_2}^4 \\
 @V h VV  @V h VV \\
 {\Z_2}^3 @> f_2 >> {\Z_2}^3
\end{CD} \textrm{ y  } \ \  \ \begin{CD}
{\Z_2}^4@> g_4 >> {\Z_2}^4 \\
 @V h VV  @V h VV \\
 {\Z_2}^3 @> f_2 >> {\Z_2}^3
\end{CD}.\]
So, $h$ is a structural  homomorphism, or an  homomorphism of PBN.
The probabilities are  $p(g_1)=p(g_3)=2/3$, and $p(g_2)=p(g_4)=1/6$,
of course the sum is 1 but this is our suggestion, because we can
have others probabilities. In fact, using the Theorem in Section 4
\cite{AA}, this network is going to similar states, that is the
boolean vectors $(0,0,0,0), \ (0,0,1,0),\ (1,0,0,0), \ (1,0,1,0),\
(0,0,0,1),$ $ \ (0,0,1,1),\ (1,0,0,1),\ (1,0,1,1)$ are going to the
state  $ (1,0,0,0)$, and  the boolean vectors $(0,1,1,0), \
(1,1,0,0),\ (0,1,0,0),\ (1,1,1,0),$ $\ (0,1,1,1), \ (1,1,0,1),\
(0,1,0,1),$ $ (1,1,1,1)$ are going to the states $ (1,1,1,0)$, and
$(1,1,1,1)$, and the equilibrium is obtained after  several
iterations of the functions, that is  the powers of the transition
matrix give the following information.
\[\pi=(0,0,.5,0,0,0,0,0,0,0,.25,0,0,0,0,.25)\]
\section{ The immune system: modeling the B/T-cells interactions}
The following model appears in \cite{P}, under the information that
it is the model $B/T-cells$, of Kaufman, Urbain, and  Thomas,
described  in 1985. Here we introduce an important applications of
the model PBN to the understanding the immune system.
\[Model \  B/T-cells\begin{array}{ccccc}
\underbrace{\overbrace{Antibody=a}}&\leftrightarrows&
\underbrace{\overbrace{e}}&\rightarrow&
\underbrace{\overbrace{T_h-cells}}\circlearrowleft\\
&\nwarrow & \downarrow & \swarrow &\uparrow\downarrow\\
& & \underbrace{\overbrace{B-cells}}& &
\underbrace{\overbrace{T_s-cells}}\circlearrowleft\\
\end{array}\]
The original model had a complex presentation, using a parameter
$e$, for the antigen, that has only two values, that is antigen is
assumed to be either present or absent and hence is represented by
$e$, a binary parameter.
 The binary variables are $b$, the B-cell
population, $a$, the secreted antibody concentration, and $h$ and
$s$, helper and suppressor T-cell populations, respectively. We use
for boolean functions, the polynomial representations over the field
$\Z_2$, then we have the following functions:
\[f_b(b,h,s,a)=eh;\ f_h(b,h,s,a)=e\overline{s}+h+e\overline{s}h;\]
\[f_s(b,h,s,a)=h+s+hs;\ f_a(b,h,s,a)=ebh.\]
that had obtained by the digraph in the above diagram, called "
Model B/T-cells". Because the parameter $e$ takes two values  0,
and 1, we obtain two functions:
\[f_0(b,h,s,a)=(0,h,h+s+hs,0); \]
\[f_1(b,h,s,a)=(h,\overline{s}+h+\overline{s}h,h+s+hs,bh)\]
Then, our space has   $2^4=16$, and we can see how the system is
moving to the  equilibrium of the system
\begin{figure}\label{fig}
\includegraphics[height = 3in,width = 4.5in]{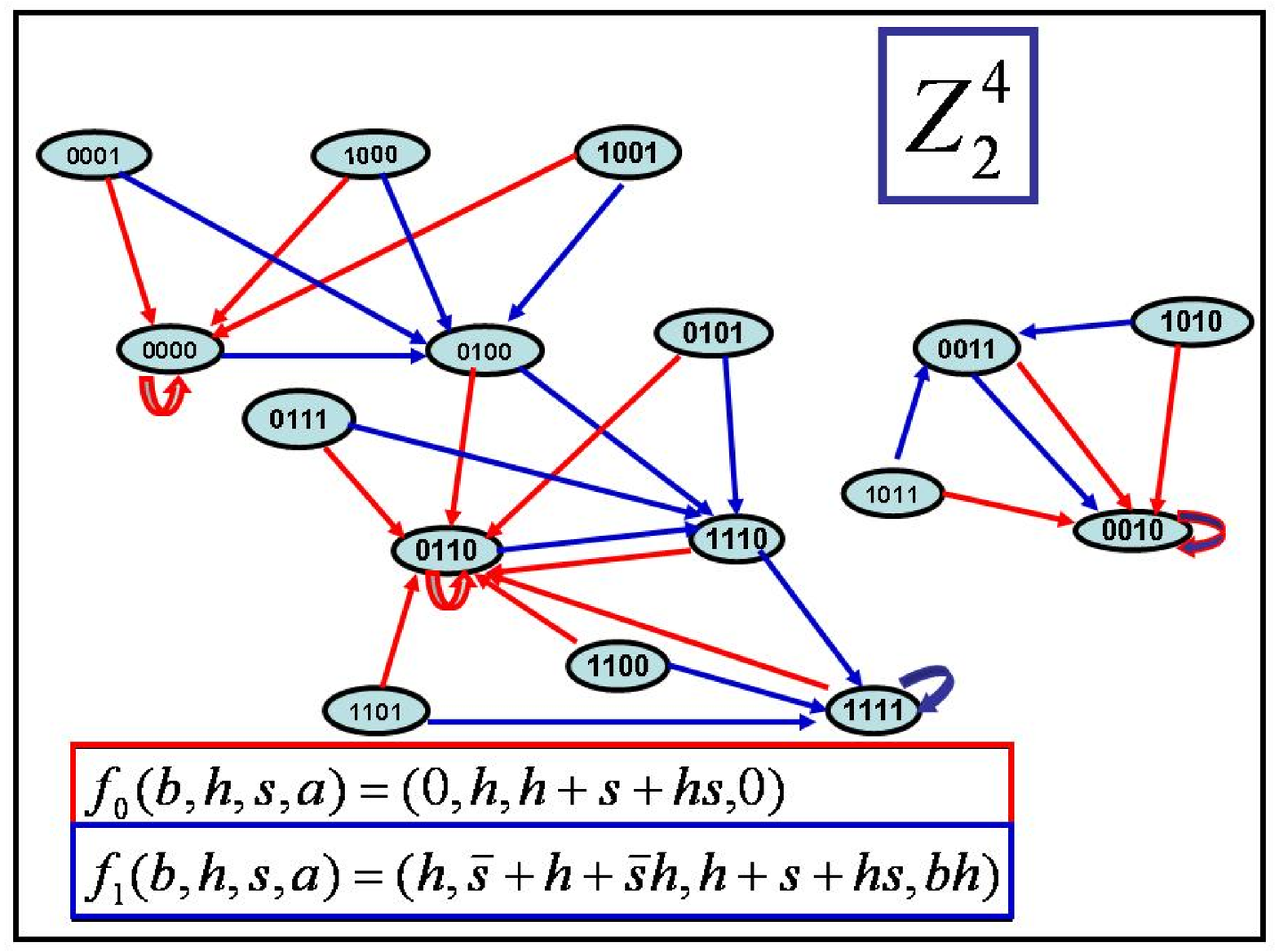}\caption{ Espacio de estados del Modelo Booleano B/T-cell}
\end{figure}
It is easy to see that there are  three steady states for the
function $f_0$, that is when the antigen is absent, and two steady
states for the function  $f_1$, that is when the antigen is present.
In \cite{P}, it is considered the function  $f_0$ working and the
same time of the function $f_1$ but they do not use probabilities,
so the biological conclusion  is not supported by a good description
of the activity in the net. They consider that the stationary points
of $f_0$ are the virgin states, that is they do not have memory
about the antigen, because in those states there are only helpers
and suppressors cells. But with our analysis we obtain a new results

Using Markov Chains, and assigning the same probability to each
function, that is .5,  the equilibrium of the system is the
following
\[\pi=(.375,0,0,0,0,0,0,0,0,0,.375,0,.125,.125,0,0),\]
where the order in the  set of states is the following $\{(0,0,1,0),
(1,0,1,1),(0,0,1,1),$\\ $(1,0,1,0),(0,0,0,0), (0,0,0,1), (0,1,0,0),
(1,0,0,0),(1,0,0,1),(0,1,1,1), (0,1,1,0),$\\
$(0,1,0,1),(1,1,1,0), (1,1,1,1), (1,1,0,0), (1,1,0,1)\}$
 We suggest the following interpretation of this equilibrium. The
 states $(0,0,1,0),$ and $(0,1,1,0)$ are the steady states when we
 do  not have antigen, and they have more probability to arrive for the network.
 Meanwhile the others two states are $(1,1,1,0)$, and $(1,1,1,1)$,
 when the interaction has a  complete action in the network, and they
 had less probabilities, because the life in general, is  going to the
 others states, maybe this happen when the helper and the suppressors
 cells are acting but they are not  enough strong to destroy the
 antigen. So a better and more complex dynamical system, can  work,
 if we consider three possibilities for the antigen, and for all
 variables in this particular model of interaction.

\end{document}